\documentclass[12pt,a4paper]{article}
\usepackage{amsfonts}
\input{epsf}
\usepackage{epsfig}
\def\IN{\mathbb{N}}
\def\IZ{\mathbb{Z}}
\def\IR{\mathbb{R}}

\begin{document}
\thispagestyle{empty}
\setcounter{page}{0}
\renewcommand{\theequation}{\thesection.\arabic{equation}}
\newcommand{\eqn}[1]{eq.(\ref{#1})}
\renewcommand{\section}[1]{\addtocounter{section}{1}
\vspace{5mm} \par \noindent
  {\bf \thesection . #1}\setcounter{subsection}{0}
  \par
   \vspace{2mm} } 
\newcommand{\sectionsub}[1]{\addtocounter{section}{1}
\vspace{5mm} \par \noindent
  {\bf \thesection . #1}\setcounter{subsection}{0}\par}
\renewcommand{\subsection}[1]{\addtocounter{subsection}{1}
\vspace{2.5mm}\par\noindent {\em \thesubsection . #1}\par
 \vspace{0.5mm} }
\renewcommand{\thebibliography}[1]{ {\vspace{5mm}\par \noindent{\bf
References}\par \vspace{2mm}}
\list
 {\arabic{enumi}.}{\settowidth\labelwidth{[#1]}\leftmargin\labelwidth
 \advance\leftmargin\labelsep\addtolength{\topsep}{-4em}
 \usecounter{enumi}}
 \def\newblock{\hskip .11em plus .33em minus .07em}
 \sloppy\clubpenalty4000\widowpenalty4000
 \sfcode`\.=1000\relax \setlength{\itemsep}{-0.4em} }
{\hfill{VUB/TENA/00/1, hep-th/0002180}}
\addtocounter{footnote}{2}

\vspace{2cm}

\begin{center}
{\bf NON-ABELIAN BORN-INFELD VERSUS STRING THEORY}

\vspace{1.4cm}

FREDERIK DENEF${}^1$, ALEXANDER SEVRIN${}^2$ and
JAN TROOST${}^2$ \footnote{Aspirant FWO, address after September 1, 2000:
CTP, MIT, Cambridge, USA.}

\vspace{.2cm}

${}^1${\em Department of Mathematics, Columbia University,}\\
{\em New York, N.Y. 10027, USA} \\
${}^2${\em Theoretische Natuurkunde, Vrije Universiteit Brussel} \\
{\em Pleinlaan 2, B-1050 Brussels, Belgium} \\
\end{center}

\vspace{-.1cm}

\centerline{{\tt denef@math.columbia.edu; asevrin,
troost@tena4.vub.ac.be}}

\vspace{1cm}

\centerline{ABSTRACT}
\vspace{- 4 mm}  
\begin{quote}\small
Motivated by the results of Hashimoto and Taylor,
we perform a detailed study of the mass spectrum of the
non-abelian Born-Infeld theory, defined by the symmetrized trace
prescription, on tori with constant magnetic fields
turned on. Subsequently, we compare this for several cases to
the mass spectrum of intersecting D-branes.
Exact agreement is found in only two cases: BPS configurations on the
four-torus and coinciding tilted branes. Finally we investigate the fluctuation
dynamics of an arbitrarily wrapped Dp-brane with flux.
\end{quote}
\baselineskip18pt
\noindent

\vspace{5mm}

\newpage
\setcounter{equation}{0}
\section{Introduction}
One of the most fascinating consequences of the discovery of D-branes is
their intimate connection with gauge theories \cite{WT1}.
The dynamics of the massless fields on an
isolated D$p$-brane is well described by a $U(1)$ Born-Infeld action.
When more than one brane is present, the situation changes. The strings
stretching between two branes have a mass proportional to the shortest
distance between those branes. When the branes coincide, additional
massless states appear. In this way, the effective action for $N$
well separated branes, described by a $(U(1))^N$ Born-Infeld theory, gets
promoted to a $U(N)$ non-abelian Born-Infeld theory \cite{EW}.
The precise form of a
non-abelian Born-Infeld theory is still unclear. The two leading terms are
known. The $F^2$ term is nothing but a $U(N)$ Yang-Mills theory, while
the $F^4$ term has been obtained directly from both the calculation of a
four point open string amplitude \cite{ssa} and from a three-loop beta-function
\cite{BP}. As the  effective
action has to reproduce tree-level open string amplitudes,
it is clear that the trace over the Lie
algebra elements should necessarily be taken outside the square root. On
this basis and from the fact that for the open superstring, odd powers of the
fieldstrength should vanish, Tseytlin formulated a conjecture for the
non-abelian Born-Infeld action \cite{AT}: before taking the trace, the Lie
algebra elements should be fully symmetrized.
Checking this proposal at the level of $F^6$ and higher directly, is a
formidable task as it requires at least either the calculation
of a six-point string amplitude or a five-loop beta-function.

In a beautiful paper
\cite{HT} (see also \cite{GR}),
it was shown that certain aspects of the full Born-Infeld
can be probed by switching on sufficiently large constant background
fields. A direct test of the non-abelian Born-Infeld theory emerged in
this way. Upon compactifying the theory on a torus and T-dualizing,
one obtains a configuration of intersecting D-branes which
allows for a string theoretic calculation of the spectrum. Subsequently,
one calculates the spectrum of the Born-Infeld theory. Consistency
requires that both results agree. However, some evidence was found
that this was not the case \cite{HT}. In \cite{B} it was shown that the
example at hand was not sufficient to uniquely resolve the discrepancy at
order $F^6$. This was the main motivation of the present work in which
tools are developed which make the calculation possible for
arbitrary configurations.

The analysis in \cite{HT} was limited to the four-torus, as only in that
case the spectrum of the off-diagonal fluctuations of the corresponding
Yang-Mills theory was known\cite{PVB}. Recently, one of us extended the
analysis of \cite{PVB} to other tori \cite{JT}. Using these results, we
systematically study the spectrum of the non-abelian Born-Infeld theory in
the presence of constant magnetic background fields on various tori. The
two-torus provides the clearest and simplest example of the discrepancy.

The outline of this paper is as follows. In the next section, we briefly
recapitulate the abelian Born-Infeld theory and its stringy
interpretation. Section three is devoted to the calculation of the quadratic
fluctuations of the non-abelian Born-Infeld theory. A general prescription
is developed which allows to perform the calculation in the presence of
arbitrary constant and abelian backgrounds. In section four we apply
these results to the simplest case: the two-torus with arbitrary magnetic
fields. We compare both Born-Infeld and string theory results.
In the following section, additional sample calculations are made: general
configurations on $T^4$ and BPS configurations on $T^6$ are studied.
The subsequent section is devoted to more involved
wrappings of Dp-branes on $T^p$.
We end with some speculations on the true structure of the
non-abelian Born-Infeld theory.

\setcounter{equation}{0}
\section{Abelian Born-Infeld}
As a warm-up exercise, we examine the spectrum of the
now well-understood abelian case and compare it to the string
theory spectrum.
The abelian Born-Infeld lagrangian describes an isolated D$p$-brane.
For a configuration with no transverse scalars excited, it is
simply\footnote{We denote space-time indices by $\alpha $, $\beta$, ... We
work in a Minkowski signature with the ``mostly plus'' convention. Space-like
indices are
denoted by $i$, $j$, ... Traces over space-time
indices are denoted by $tr$, while a trace over groupvariables is written as
$Tr$. A
symmetrized trace over groupvariables is denoted by $STr$. We work in units
where $2\pi\alpha '=1$ and we ignore an overall factor.}
\begin{eqnarray}
{\cal L} &=& - \, \left(\det
(\delta_{\alpha}^{\, \beta}+ 2 \pi \alpha' F_{\alpha}^{\, \beta})
\right)^{\frac{1}{2}}, \label{abi}
\end{eqnarray}
with $F_{\alpha}{}^{\beta}\equiv\partial_\alpha
A^\beta-\partial^\beta A_\alpha $.
We expand the gauge field $A_\alpha $ around a fixed background
${\cal A}_\alpha $, $A_\alpha = {\cal A}_\alpha  +\delta A_\alpha $ and
choose the background such that its fieldstrength ${\cal F}$ is constant.
Eq. (\ref{abi}) becomes,
\begin{eqnarray}
{\cal L} &=& - \Big( \det(1+{\cal F}+\delta F) \Big)^{\frac{1}{2}}, \\
         &=& -  (\det {\cal G})^{-\frac{1}{4}}\Big(
         \det(1+{\cal G} \delta F+{\cal B} \delta F)\Big)^{\frac{1}{2}} ,
\end{eqnarray}
where
\begin{eqnarray}
{\cal G} &\equiv& (1-{\cal F}^2)^{-1}, \label{defg}\\
{\cal B} &\equiv& -{\cal F} (1-{\cal F}^2)^{-1}.\label{defb}
\end{eqnarray}
We ignore the leading constant term and drop the term linear in the
fluctuations as it is a total derivative. The first non-trivial
term is quadratic in the fluctuations and upon partial integration it
assumes a simple form,
\begin{eqnarray}
{\cal L}_{quad} = \frac 1 4  (\det {\cal G})^{-\frac{1}{4}} tr ({\cal G} \delta
F
{\cal G} \delta F ).
\end{eqnarray}
When only magnetic background fields are turned on, i.e. ${\cal F}_{0i}=0$,
this reduces to
\begin{eqnarray}
{\cal L}_{quad} &=& \frac 1 2 (\det {\cal G})^{-\frac{1}{4}}
\Big(  {\cal G}^{ij} \delta F_{0i} \delta F_{0j}
-\frac{1}{2} {\cal G}^{ij}  {\cal G}^{kl}
\delta F_{ik}  \delta F_{jl}  \Big).
\end{eqnarray}
By an orthogonal transformation, we can always bring the background fields
in a canonical form such that only ${\cal F}_{12}$, ${\cal F}_{34}$,
${\cal F}_{56}$, ... are possibly different from zero. Doing this and
introducing
$\lambda_i$,
\begin{eqnarray}
\lambda_{2i-1}=\lambda_{2i}\equiv1+({\cal F}_{2i-1\, 2i})^2,
\end{eqnarray}
we obtain the final form for the lagrangian,
\begin{eqnarray}
{\cal L}_{quad} &=& \frac 1 2 \sqrt{\prod_{k}\lambda_k}\Big(
 \lambda^{-1}_i \delta F_{0i} \delta F_{0i}
-\frac{1}{2} \lambda_i^{-2}
\delta F_{ij}  \delta F_{ij}  \Big). \label{abi1}
\end{eqnarray}
We compactify some of the spacelike directions on
a torus, and turn on magnetic fields in these directions
only. We subsequently determine the spectrum of masses in the
uncompactified directions\footnote{We use the same convention for
general spacelike indices and those parametrizing
the torus. The context should make it clear what is meant.}.
We take the compactified space to be $T^{2n}$ and write the length
of the $i^{th}$ cycle, ${\cal C}_i$, as $L_i$.
Upon rescaling the space-like directions, eq. (\ref{abi1}) becomes, modulo an
irrelevant
overall scale factor, an ordinary abelian Yang-Mills theory. In this way we
immediately obtain the spectrum of the abelian Born-Infeld theory,
\begin{eqnarray}
M^2=   \sum_i \frac {1}{\lambda_i}\Big(\frac{2 \pi m_i}
{L_i}\Big)^2,\label{sp1}
\end{eqnarray}
where the $m_i$ are integers.

We now make contact with string theory \cite{HT}. In order that the transition
functions are well defined, the first Chern class should be an integer.
In order to achieve this, we parametrize the background as,
\begin{eqnarray}
{\cal F}_{2i-1\,2i}=\frac{2\pi n_i}{L_{2i-1}L_{2i}}, \quad
{\cal A}_{2i-1}=0,\quad {\cal A}_{2i}=
{\cal F}_{2i-1\,2i}x^{2i-1},\label{fs}
\end{eqnarray}
where $n_i$ are integers.
This configuration corresponds to a single D$2n$-brane wrapped
around $T^{2n}$ with a certain number of lower dimensional D-branes
dissolved in it.
To calculate its spectrum from string theory we closely follow \cite{HT}
and T-dualize along the ${\cal C}_{2i}$, $i\in\{1,\cdots,n\}$ cycles.
The resulting configuration is
a single D$n$-brane wrapped once around the ${\cal C}'{}_{2i-1}$ cycles and
$n_i$ times around the ${\cal C}'{}_{2i}$ cycles of the dual torus.
The sizes of the cycles of the dual torus are
$L'_{2i-1}=L_{2i-1}$ and $L'_{2i}=2\pi L_{2i}^{-1}$.
Calculating the spectrum of the dual configuration
using string theory is now straightforward and yields
\begin{eqnarray}
M^2= \sum_{i=1}^n \bigg(\frac{4\pi^2m'{}_{2i-1}^2}
{L'{}^2_{2i-1}+n_i^2L'{}^2_{2i}}
+\frac{m'{}^2_{2i}L'{}^2_{2i-1}n_i^2L'{}^2_{2i}}{L'{}^2_{2i-1}+
n_i^2L'{}^2_{2i}}\bigg). \label{sp2}
\end{eqnarray}
Identifying the integers $m'_{2i-1}$ and $n_im'_{2i}$ with $m_{2i-1}$
and $m_{2i}$ respectively and switching back to the original variables,
shows that eqs. (\ref{sp1}) and (\ref{sp2}) match. In other words, both
string theory and the Born-Infeld theory yield the same spectrum.

\setcounter{equation}{0}
\section{Quadratic fluctuations of non-abelian Born-Infeld}
Turning to the non-abelian theory, we introduce Lie algebra valued
gauge fields $A_\alpha $ and their fieldstrength
$F_{\alpha \beta}= \partial_\alpha A_\beta- \partial_\beta A_\alpha - i
{[}A_\alpha ,A_\beta {]}$.
Following \cite{AT}, we define the non-abelian Born-Infeld lagrangian by
\begin{eqnarray}
{\cal L} &=& - STr \left(\det
(\delta_{\alpha}^{\, \beta}+  F_{\alpha}^{\, \beta})
\right)^{\frac{1}{2}}.
\end{eqnarray}
Taking a symmetrized trace means that upon expanding the action in powers
of the fieldstrength, one first symmetrizes all the Lie algebraic factors
and subsequently one takes the trace.

We consider a constant background fieldstrength ${\cal F}$ and denote the
background connection by ${\cal A}$. All background fields take values
in the Cartan subalgebra (CSA). We parametrize the fieldstrength as $F={\cal
F}+\delta F$
and the connection by $A={\cal A}+\delta A$.
In terms of these variables, the lagrangian becomes,
\begin{eqnarray}
{\cal L} &=& -STr \left( \det(1+{\cal F}+\delta F) \right)^{\frac{1}{2}} \\
         &=& -STr  (\det {\cal G})^{-\frac{1}{4}} \Big(
         \det(1+{\cal G} \delta F+{\cal B} \delta F)
         \Big)^{\frac{1}{2}},
\end{eqnarray}
where ${\cal G}$ and ${\cal B}$ were defined in eqs. (\ref{defg}) and
(\ref{defb}). Picking out the term quadratic in the fluctuations gives
\begin{eqnarray}
{\cal L}_{quad} &=& - STr \Big( (\det {\cal G})^{-\frac{1}{4}}
(\frac{1}{2} tr {\cal B} \delta_2 F -\frac{1}{4} tr ({\cal G} \delta_1 F
{\cal G} \delta_1 F\nonumber \\
& &  + {\cal B} \delta_1 F  {\cal B} \delta_1 F)
+\frac{1}{8} (tr {\cal B} \delta_1 F)^2) \Big),\label{kwad}
\end{eqnarray}
where we used,
\begin{eqnarray}
D \cdot&\equiv& \partial - i {[} {\cal A} , \cdot {]}, \\
\delta F &\equiv& \delta_1 F + \delta_2 F, \\
\delta_1 F_{\alpha \beta} &\equiv& D_{\alpha} \delta A_{\beta}
                                   -D_{\beta} \delta A_{\alpha} ,\\
\delta_2 F_{\alpha \beta} &\equiv& -i {[}\delta A_{\alpha}, \delta A_{\beta}
{]} .
\end{eqnarray}

Again, we take the background field purely magnetic, implying ${\cal
G}^{00}=-1,
{\cal G}^{0i}= {\cal B}^{0i} = 0, {\cal B}^{ij}=- ({\cal F})^i_k
{\cal G}^{kj}$. The lagrangian reads then,
\begin{eqnarray}
 {\cal L}_{quad} &=& - STr \Big( (\det {\cal G})^{-\frac{1}{4}}
 ( -\frac{1}{2}  {\cal B}^{ij} \delta_2 F_{ij}
 -\frac{1}{2}  {\cal G}^{ij} \delta_1 F_{0i} \delta_1 F_{0j}\nonumber\\
&&+\frac{1}{4} {\cal G}^{ij}  {\cal G}^{kl}
\delta_1 F_{ik}  \delta_1 F_{jl}     +    \frac{1}{4}
{\cal B}^{ij}  {\cal B}^{kl} \delta_1 F_{il}  \delta_1 F_{jk})
+\frac{1}{8}  ({\cal B}^{ij} \delta_1 F_{ij})^2) \Big). \label{int1}
\end{eqnarray}
The terms proportional to ${\cal B}^2$ lead, upon partial integration, to,
\begin{eqnarray}
-\frac{3i}{4} Str(\det{\cal G})^{-\frac 1 4}{\cal B}^{i j}
{\cal B}^{kl}{[}{\cal F}_{{[}ij},\delta A_{k{]}}{]}
\delta A_l, \label{BB}
\end{eqnarray}
where, when performing the symmetrized trace, the commutator term has to
be considered as a single Lie algebra element. Later on, we will see
that this term gives an
additional contribution to the zero point energy.

As an intermezzo, we now turn to the calculation of the symmetrized traces.
In the analysis of the lagrangian at the quadratic level, we get two kinds of
contributions: those linear and those quadratic in the field strength
variation.
Recalling that the background field strength takes values in the CSA, we get
that for the
linear terms, the symmetrized trace reduces to an ordinary trace.
\begin{eqnarray}
\mbox{type 1}\equiv STr ( \delta_2 F {\cal F}_1 {\cal F}_2 \cdots {\cal
F}_{n-1} )&=&
Tr (\delta_2 F   {\cal F}_1 {\cal F}_2 \cdots {\cal F}_{n-1})
\end{eqnarray}
We turn now to the
contributions quadratic in the variation of the field strength. They are
of the form,
\begin{eqnarray}
\mbox{type 2} &\equiv& STr (\delta_1 \bar{F} \delta_1 F
{\cal F}_1 {\cal F}_2 \cdots {\cal F}_{2n})\nonumber\\
&&= \frac{1}{(2n+1)!} Tr \{\delta_1 \bar{F}
(\delta_1 F {\cal F}_1 {\cal F}_2 \cdots {\cal F}_{2n} + \mbox{all
permutations}) \} ,
\end{eqnarray}
where the presence of the subindices and the bar above the first variation
reflects the possibility that these various terms might have
different space-time index structures.
Since all fields are in
the fundamental representation of the group $U(N)$, \cite{AT}, we
have
\begin{eqnarray}
{\cal F}&=&\sum_{a=1}^N {\cal F}^a E_{aa},\nonumber\\
\delta F&=& \sum_{a,b=1}^N \delta F^{ab} E_{ab}, \label{Eab}
\end{eqnarray}
where $E_{ab}$  is an $N\times N$ matrix unit, i.e. it is an
$N\times N$ matrix which is zero except on the $a^{th}$ row and $b^{th}$
column where there is a 1. Using this notation, we can perform the trace
in the type 2 term,
\begin{eqnarray}
\mbox{type 2} & = &\frac{1}{(2n+1)!}\sum_{a,b=1}^N \delta_1
\bar{F}^{ba}\delta_1 F^{ab}
\nonumber \\
& & \Big(2n!   {\cal F}_1^b {\cal F}_2^b \cdots {\cal F}_{2n}^b  + (2n-1)!
{\cal F}_1^a
 {\cal F}_2^b \cdots {\cal F}_{2n}^b + (2n-1)! {\cal F}_1^b {\cal F}_2^a {\cal
F}_3^b
 \cdots {\cal F}_{2n}^b\nonumber\\
&&+\cdots + (2n-1)! {\cal F}_1^b
 {\cal F}_2^b \cdots {\cal F}_{2n-1}^b {\cal F}_{2n}^a
+ 2! (2n-2)!  {\cal F}_1^a  {\cal F}_2^a   {\cal F}_3^b \cdots  {\cal F}_{2n}^b
  +
\cdots \nonumber \\
& & \cdots + 2n!   {\cal F}_1^a  {\cal F}_2^a  \cdots {\cal F}_{2n}^a  \Big) .
\end{eqnarray}
{}From this expression we see that we can look at each sector with given
$a$ and $b$ separately. In other words we work with a $U(2)$ subgroup.
Defining ${\cal F}^a = {\cal F}^0 + {\cal F}^3$ and
${\cal F}^b = {\cal F}^0 - {\cal F}^3$, we get for fixed $a$ and $b$,
\begin{eqnarray}
\mbox{type 2}
&=& \Big(\delta_1 \bar{F}^{ba} \delta_1 F^{ab}+\delta_1 \bar{F}^{ab} \delta_1
F^{ba}
\Big)
\Big( {\cal F}_1^0  \cdots {\cal F}_{2n}^0  +\frac{1}{3}
(   {\cal F}_1^3 {\cal F}_2^3 {\cal F}_3^0 \cdots {\cal
F}_{2n}^0\nonumber\\
&&+  {\cal F}_1^3 {\cal F}_2^0 {\cal F}_3^3   {\cal F}_4^0
\cdots {\cal F}_{2n}^0 +\cdots +
{\cal F}_1^0\cdots {\cal F}_{2n-2}^0 {\cal F}_{2n-1}^3 {\cal
F}_{2n}^3) \nonumber \\
& & + \frac{1}{5} ({\cal F}_1^3{\cal F}_2^3{\cal F}_3^3{\cal F}_4^3
{\cal F}_5^0\cdots{\cal F}_{2n}^0+\cdots)+\cdots \nonumber \\
& & \cdots + \frac{1}{2n+1}  {\cal F}_1^3  \cdots {\cal F}_{2n}^3
\big).
\end{eqnarray}
This suggests a simple way to implement the symmetric trace. Given some
arbitrary even function of the backgroundfields $H({\cal F})$, we get in a
$U(2)$ subsector,
\begin{eqnarray}
STr (\delta_1 \bar{F} \delta_1 F H({\cal F}))=
\Big(\delta_1 \bar{F}^{ba} \delta_1 F^{ab}+\delta_1 \bar{F}^{ab} \delta_1
F^{ba}
\Big){\cal I}(H), \label{pres}
\end{eqnarray}
where
\begin{eqnarray}
{\cal I}(H)\equiv\frac 1 2  \int^1_0 d\alpha
\Big( H({\cal F}^0 + \alpha {\cal F}^3)+
H({\cal F}^0 -\alpha {\cal F}^3\Big).\label{pres1}
\end{eqnarray}

These results allow for the calculation of the symmetrized trace through
second order in the fluctuations in the presence of constant abelian
background. In the remainder of this section, we will use this to
make the action eq. (\ref{int1}) as explicit as
possible. We consider the Born-Infeld theory on an even-dimensional torus
$T^{2n}$.
This time an orthogonal transformation is not sufficient to bring the
background into a form where only $f_i\equiv {\cal F}_{2i-1\,2i}$,
for $i\in\{1,2,\dots , n\}$, are non-zero. In order not to
overload the formulae, we will assume this anyway.
Furthermore, without loss of
any generality, we will work with a $U(2)$ theory. The background is
of the form
\begin{eqnarray}
f_i=f_i^0\sigma_0 + f_i^3\sigma_3,
\end{eqnarray}
where $\sigma_0$ is the $2\times 2$ unit matrix and $\sigma_1$, $\sigma_2$
and $\sigma_3$ are the Pauli matrices.
The fluctuations are written as
\begin{eqnarray}
\delta F = \sum_{a=0}^3\delta F^{(a)}\sigma_a.
\end{eqnarray}

Combining the term proportional to $\delta_2{\cal F}$ with eq. (\ref{BB}),
we get the zero-point energy term\footnote{Note that we slightly abuse
language as also part of the potential term will
contribute to the zero-point energy.},
\begin{eqnarray}
{\cal L}_0&=&\sum_{i=1}^n\delta F^{(3)}_{2i-1\,2i}\Bigg\{ -
\sqrt{\frac{\prod_{k\neq i} (1+(f_k^0+f_k^3)^2)}{1+(f_i^0+f_i^3)^2}}
(f^0_i+f^3_i)+\nonumber\\
&&\qquad\sqrt{\frac{\prod_{k\neq i}
(1+(f_k^0-f_k^3)^2)}{(1+(f_i^0-f_i^3)^2}}(f^0_i-f^3_i)+\nonumber\\
&&\qquad 2\sum_{j\neq i}f_j^-{\cal I} \bigg(f_if_j \sqrt{\frac{\prod_{k\neq
i,j}(
1+f_k^2 ) }{(1+f_i^2)(1+f_j^2) }} \ \bigg)\Bigg\}. \label{l1}
\end{eqnarray}
In a similar way, we obtain from eq. (\ref{int1}) the kinetic term
\begin{eqnarray}
{\cal L}_1= \sum_{i=1}^n{\cal I}\Bigg\{ \sqrt{\frac{\prod_{k\neq i}(
1+f_k^2 ) }{1+f_i^2 }}\ \Bigg\} \sum_{a=1}^2\Big( (\delta_1F^{(a)}_{0\,
2i-1})^2 +(\delta_1 F^{(a)}_{0\, 2i})^2\Big),\label{l2}
\end{eqnarray}
and the potential
\begin{eqnarray}
{\cal L}_2&=& -\frac 1 2 \sum_{i=1}^n\sum_{j\neq i}
{\cal I}\Bigg\{ \sqrt{\frac{\prod_{k\neq i,j}(
1+f_k^2 ) }{(1+f_i^2)(1+f_j^2) }}\ \Bigg\} \sum_{a=1}^2\Big(
(\delta_1F^{(a)}_{2i-1\,
2j-1})^2 +\nonumber\\
&&\qquad (\delta_1 F^{(a)}_{2i\, 2j})^2 +2(\delta_1 F^{(a)}_{2i-1\,
2j})^2\Big)-\nonumber\\
&&\sum_{i=1}^n
{\cal I}\Bigg\{ \sqrt{\frac{\prod_{k\neq i}(
1+f_k^2 ) }{(1+f_i^2)^3 }}\ \Bigg\} \sum_{a=1}^2\Big(
(\delta_1 F^{(a)}_{2i-1\, 2i})^2\Big).\label{l3}
\end{eqnarray}
The total Born-Infeld lagrangian is the sum of ${\cal L}_0$,
${\cal L}_1$ and ${\cal L}_2$.
Using gauge invariance, one can show that the rescaling in eq. (\ref{l1})
coincides with the last rescaling in eq. (\ref{l3}).
As the expression for the lagrangian is
rather involved, we turn in the next two sections to some specific
examples which we then compare to string theory.

\setcounter{equation}{0}
\section{A case study: D2-branes on $T^2$ }
The simplest case at hand are two $D2$-branes on $T^2$. T-dualizing along
one of the directions gives two D1-branes intersecting each other at a
certain angle. This picture allows for a string theoretic calculation of the
spectrum. Subsequently, this can be compared to the spectrum obtained from
the non-abelian Born-Infeld theory.

\subsection{The spectrum from string theory}
We consider IIB string theory on $T^2\times M_8$, where $M_8$ is the
eight-dimensional Minkowski space, in the presence of two D1-branes
wrapped around $T^2$ and intersecting each other at an angle $\gamma$.
We take the two branes along the 1-axis and rotate one of them
over an angle $\gamma$ into the 12-plane.
We now proceed with the calculation of the
masses low-lying states thereby closely following \cite{BDL} (see also
\cite{CC} and \cite{CP}).

Combining the string coordinates on the torus as
$Z\equiv X^1+i X^2$, we get the boundary conditions for the end point
of a string tied to the first D1-brane
\begin{eqnarray}
\mbox{Im} \,Z|_{\sigma=0}&=&0,\nonumber\\
\mbox{Re}\,\partial_\sigma Z|_{\sigma=0}&=&0,
\end{eqnarray}
and for the end point tied to the rotaded brane
\begin{eqnarray}
\mbox{Im}\, e^{i\gamma}Z|_{\sigma=\pi}&=&0,\nonumber\\
\mbox{Re}\,\partial_\sigma e^{i\gamma} Z|_{\sigma=\pi}&=&0.
\end{eqnarray}
Implementation of the boundary conditions leads to the following mode
expansion
\begin{eqnarray}
Z=\sum_{m\in\IZ}\left(a_{m-\beta}e^{i(m-\beta)(\tau+\sigma)}+
a_{-m-\beta}^\dagger e^{i(m+\beta)(\tau-\sigma)}\right),
\end{eqnarray}
where $\beta=\gamma/\pi$.
Using,
\begin{eqnarray}
\frac{d}{dx}\frac{e^{a x}}{1-e^x}&=&\frac{1}{x^2}+ \frac{1}{12}(-6(a-1)a-1)
-\frac{1}{6}((a-1)a(2a-1)) x +{\cal O}(x^2),\nonumber\\
\frac{d}{dx}\frac{e^{(1+a) x}}{1-e^{2x}}&=&\frac{1}{2x^2}+ \frac{1}{12}(1-
3a^2)
+\frac{1}{6}(a-a^3) x +{\cal O}(x^2),
\end{eqnarray}
we get the regularized expressions for the vacuum energy
of a boson with a moding shifted by $\beta$,
\begin{eqnarray}
\frac 1 2 \sum_{n=1}^\infty (n-\beta)=-\frac{1}{24} + \frac 1 4
\beta(1-\beta),
\end{eqnarray}
and for Ramond and Neveu-Schwarz fermions respectively,
\begin{eqnarray}
-\frac 1 2 \sum_{n=1}^\infty (n-\beta)&=&\frac{1}{24} - \frac 1 4
\beta(1-\beta),\nonumber\\
-\frac 1 2 \sum_{n=1}^\infty (n-\frac 1 2 -\beta)&=&-\frac{1}{48} + \frac 1 4
\beta^2.
\end{eqnarray}
For the configuration under study, we find that the vacuum energy in the
Ramond sector vanishes,
while in the Neveu-Schwarz sector it is given by:
\begin{eqnarray}
&&2\pi\Big(6\times (-\frac{1}{24}) + 2\times (-\frac{1}{24}+\frac 1 4
\beta(1-\beta))+ \nonumber\\
&&\qquad 6\times (-\frac{1}{48})  + 2\times ( -\frac{1}{48} + \frac 1 4
\beta^2  )\Big)=\gamma-\pi.
\end{eqnarray}
Now it becomes simple to calculate the masses of the low-lying states in
the Neveu-Schwarz sector. The relevant states are
\begin{eqnarray}
(a_{-\beta})^m\psi_{-\frac 1 2 +\beta}|0>_{NS}
\end{eqnarray}
with a mass given by
\begin{eqnarray}
M^2&=&(2m-1)\gamma ,\label{res1a}
\end{eqnarray}
and
\begin{eqnarray}
(a_{-\beta})^m\psi'{}_{-\frac 1 2 -\beta}|0>_{NS}
\end{eqnarray}
with a mass given by
\begin{eqnarray}
M^2&=&(2m+3)\gamma .\label{res1b}
\end{eqnarray}

\subsection{The spectrum from Born-Infeld theory}
We will restrict ourselves to the study of the off-diagonal gauge field
fluctuations. This can easily be generalized, without altering our results,
to include the transverse scalars and the fermions as well.
We consider the $U(2)$ Born-Infeld lagrangian on $T^2$ and choose the
diagonal background
\begin{eqnarray}
f\equiv{\cal F}_{12}=f^0 \sigma_0 + f^3 \sigma_3.
\end{eqnarray}
{}From eqs. (\ref{l1}-\ref{l2}), we get an expression for the part of the
Lagrangian quadratic in the fluctuations,
\begin{eqnarray}
{\cal L}_{quad} &=& \Bigg\{ -\frac{f^0+f^3}{\sqrt{1+(f^0+f^3)^2}}
+\frac{f^0-f^3}{\sqrt{1+(f^0-f^3)^2}}\Bigg\}\delta_2F_{12}^{(3)}+\nonumber\\
&&{\cal I}\bigg\{\frac{1}{\sqrt{1+f^2}}\bigg\}\sum_{a=1}^2\sum_{i=1}^2
\Big(\delta_1F_{0i}^{(a)}\Big)^2-
{\cal I}\bigg\{\frac{1}{\sqrt{(1+f^2)^3}}\bigg\}\sum_{a=1}^2
\Big(\delta_1F_{12}^{(a)}\Big)^2. \nonumber\\
\end{eqnarray}
Performing the symmetrized trace integrals, we obtain the final result,
\begin{eqnarray}
{\cal L}_{quad} &=& -\left\{ \frac{f^0+f^3}{\sqrt{1+(f^0+f^3)^2}}-
\frac{f^0-f^3}{\sqrt{1+(f^0-f^3)^2}}\right\}\delta_2 F_{12}^{(3)}+\nonumber\\
&&\frac{\mbox{arcsinh}(f^0+f^3)-\mbox{arcsinh}(f^0-f^3) }{2f^3}
\sum_{i,a=1}^2\delta_1 \Big( F_{0i}^{(a)}\Big)^2-\nonumber\\
&&\frac{1}{2f^3} \left\{ \frac{f^0+f^3}{\sqrt{1+(f^0+f^3)^2}}-
\frac{f^0-f^3}{\sqrt{1+(f^0-f^3)^2}}\right\} \sum_{a=1}^2
\Big( \delta_1  F_{12}^{(a)}\Big)^2.
\end{eqnarray}

{}From this it follows that the spectrum of the non-abelian Born-Infeld
theory is
the same as that of the corresponding non-abelian Yang-Mills theory, but
rescaled by a factor $\varepsilon$,
\begin{eqnarray}
\varepsilon&=& \left\{ \frac{f^0+f^3}{\sqrt{1+(f^0+f^3)^2}}-
\frac{f^0-f^3}
{\sqrt{1+(f^0-f^3)^2}}\right\}\times\nonumber\\
&&\qquad\frac{1}
{\mbox{arcsinh}(f^0+f^3)-\mbox{arcsinh}(f^0-f^3)}.
\end{eqnarray}
The spectrum of a $U(2)$ Yang-Mills theory on $T^{2n}$ was calculated in
\cite{JT}.
Combining this with the previous, we get the Born-Infeld spectrum for the
off-diagonal fluctuations of the gauge fields, it is given by
\begin{eqnarray}
M^2&=&2(2m-1)\varepsilon f^3,\nonumber\\
M^2&=&2(2m+3)\varepsilon f^3. \label{res2}
\end{eqnarray}

\subsection{Comparing results}
T-dualizing the Born-Infeld configuration, we get two D1-branes wrapped
around $T^2$ and intersecting each other at an angle $\gamma$ given by
\begin{eqnarray}
\gamma= \mbox{arctan}\left\{ \frac{2f^3}{1+(f^0)^2
-(f^3)^2}\right\}.\label{res3}
\end{eqnarray}
Combining eqs. (\ref{res1a}) and (\ref{res1b}) with eq. (\ref{res3}) and
comparing it to eq. (\ref{res2}), shows that the spectrum calculated from
string theory does not match the spectrum predicted by the non-abelian
Born-Infeld theory. Of course, in the limit when $f^3$ vanishes, which
probes the abelian sector, both results match. The Taylor expansion
of the Born-Infeld rescaling around $f^0=f^3=0$ consists of
terms of the form $(f^0)^{2m}(f^3)^{2n}$, $m,\,n\in\IN$.
Such a term arises from the $F^{2m+2n+2}$ terms in
the non-abelian Born-Infeld action. Agreement with the
string theoretic results occurs only for
$(m,n)=(m,0)$ and $(m,n)=(0,1)$, from which it follows that through order
$F^4$, the Born-Infeld
action gives correct results. The mismatch appearing at order $F^6$
and higher is quite serious.

\setcounter{equation}{0}
\section{Some other cases}

\subsection{D4 on $T^4$}
In \cite{HT}, $D4$ configurations on $T^4$ were investigated. We reexamine
this case and show that for self-dual configurations, the string theory
results agree with the non-abelian Born-Infeld predictions. However, we
provide strong indications that any
other configuration will disagree at order $F^6$ and higher.

We start by taking two 2-branes on $T^4$ along the 14 plane and rotating one
of them first over an angle $\gamma_1$ in the 12 plane and subsequently
over an angle $\gamma_2$ in the 43 plane. The spectrum was obtained in
\cite{HT}. The low-lying NS states have masses
\begin{eqnarray}
M^2&=&(2m_1-1)\gamma_1 +
(2 m_2+1 )\gamma_2,\nonumber\\
M^2&=&(2m_1+1)\gamma_1 +
(2 m_2-1 )\gamma_2,\nonumber\\
M^2&=&(2 m_1+3 ) \gamma_1 +
(2 m_2+1 )\gamma_2,\nonumber\\
M^2&=&(2 m_1+ 1 )\gamma_1 +
(2m_2+ 3 )\gamma_2, \label{d4}
\end{eqnarray}
with $m_1$ and $m_2$, two integers.

We choose a simple background
\begin{eqnarray}
{\cal F}_{12}=f_1\sigma_3,\qquad {\cal F}_{34}=f_2\sigma_3.
\end{eqnarray}
The background fields are related to the angles by
\begin{eqnarray}
\tan \frac {\gamma_1}{2}=f_1,\qquad
\tan \frac {\gamma_2}{2}=f_2. \label{angles}
\end{eqnarray}
Eqs. (\ref{l1}-\ref{l2}) yield
\begin{eqnarray}
{\cal L}_{quad} &=& \Bigg\{ -2 f_1\sqrt{\frac{ 1+f_2^2}{1+f_1^2}}
+ 2 f_2{\cal I}\bigg( \frac{f_1f_2}{\sqrt{(1+f_1^2)(1+f_2^2)}}
\bigg)\Bigg\}\delta_2F_{12}^{(3)}+ \nonumber\\
&&\Bigg\{ -2 f_2\sqrt{\frac{ 1+f_1^2}{1+f_2^2}}
+ 2 f_1{\cal I}\bigg( \frac{f_1f_2}{\sqrt{(1+f_1^2)(1+f_2^2)}}
\bigg)\Bigg\}\delta_2F_{34}^{(3)} +\nonumber\\
&&{\cal I}\bigg\{\sqrt{\frac{1+f_2^2}{1+f_1^2}}\bigg\}\sum_{a=1}^2\sum_{i=1}^2
\Big(\delta_1F_{0i}^{(a)}\Big)^2+
{\cal I}\bigg\{\sqrt{\frac{1+f_1^2}{1+f_2^2}}\bigg\}\sum_{a=1}^2\sum_{i=3}^4
\Big(\delta_1F_{0i}^{(a)}\Big)^2 -\nonumber\\
&&{\cal I}\bigg\{\sqrt{\frac{1+f_2^2}{(1+f_1^2)^3}}\bigg\}\sum_{a=1}^2
\Big(\delta_1F_{12}^{(a)}\Big)^2-
{\cal I}\bigg\{\sqrt{\frac{1+f_1^2}{(1+f_2^2)^3}}\bigg\}\sum_{a=1}^2
\Big(\delta_1F_{34}^{(a)}\Big)^2- \nonumber\\
&&{\cal I}\bigg\{\frac{1}{\sqrt{(1+f_1^2)(1+f_2^2)}}\bigg\}\sum_{a=1}^2\bigg(
\Big(\delta_1F_{13}^{(a)}\Big)^2
+\Big(\delta_1F_{24}^{(a)}\Big)^2+\nonumber\\
&&\Big(\delta_1F_{14}^{(a)}\Big)^2
+\Big(\delta_1F_{23}^{(a)}\Big)^2\bigg).\label{t4}
\end{eqnarray}
The simplest configuration is the self-dual one $f\equiv f_1=f_2$, which
corresponds
to a BPS configuration \cite{BDL}.
In that case, the integrals are standard and we get
\begin{eqnarray}
{\cal L}_{quad} &=& -2\,\frac{\mbox{arctan} (f)}{f} f
\Big(\delta_2F_{12}^{(3)}+ \delta_2F_{34}^{(3)}\Big) +
\sum_{a=1}^2\sum_{i=1}^4
\Big(\delta_1F_{0i}^{(a)}\Big)^2- \nonumber\\
&& \frac 1 2 \frac{\mbox{arctan} (f)}{f} \sum_{a=1}^2 \sum_{i,j=1}^4
\Big(\delta_1 F_{ij}^{(a)}\Big)^2.
\end{eqnarray}
So the spectrum of the off-diagonal fluctuations of
the non-abelian Born-Infeld theory is that of the
corresponding Yang-Mills theory, but rescaled by a factor $\varepsilon$,
\begin{eqnarray}
\varepsilon = \frac{\mbox{arctan}(f)}{f}=\frac{\gamma}{2f}.
\end{eqnarray}
In \cite{PVB}, one finds the spectrum of the $U(2)$ Yang-Mills theory
on $T^4$. It is given by
\begin{eqnarray}
M^2&=&2(2m_1-1)f_1 +
2(2 m_2+1 )f_2,\nonumber\\
M^2&=&2(2m_1+1)f_1 +
2(2 m_2-1 )f_2,\nonumber\\
M^2&=&2(2 m_1+3 ) f_1 +
2(2 m_2+1 )f_2,\nonumber\\
M^2&=&2(2 m_1+ 1 )f_1 +
2(2m_2+ 3 )f_2. \label{mym}
\end{eqnarray}
Rescaling this spectrum by $\varepsilon$ and comparing it to the string
spectrum, eq. (\ref{d4}), shows that for self-dual configurations,
$f\equiv f_1=f_2$ or $\gamma=\gamma_1=\gamma_2$,
there is perfect agreement! The apparent
disagreement found in \cite{HT} was due to the fact that the contribution
of the ${\cal B}^2$ terms to the zero point energy, eq. (\ref{BB}), was not
properly taken into account.

We now turn to the more complicated case, where $f_1\neq f_2$. This time
the integrals are more involved and can only be expressed in terms of
elliptic integrals of the first and second kind. This can be circumvented
by Taylor expanding the coefficients in eq. (\ref{t4}) around $f_1=f_2=0$.
However, a more serious problem arises here as well:
no linear coordinate transformation can bring eq.
(\ref{t4}) into a pure Yang-Mills form. In order to calculate the
spectrum, one could repeat the analysis of \cite{JT} which will be
complicated by the fact that the mass operator will not be diagonal in the
Lorentz indices. However we believe that it is highly unlikely that the
Born-Infeld action would reproduce the string spectrum. In fact this can
already be seen by putting the $f_2$ background to zero. For this choice,
the calculation reduces to the one studied in the previous section where no
agreement was found.

Keeping only those corrections induced by the $F^2$ and $F^4$ terms
in the Born-Infeld action gives a spectrum as in eq. (\ref{mym}), but with
$f_i$ replaced by $f_i-(f_i)^3/3$. Comparing this to eq. (\ref{d4}), using
eq. (\ref{angles}), shows that this is indeed the desired result up to
this order.

\subsection{BPS configurations on $T^6$}
{}From the previous examples, one would be tempted to conclude that
Tseytlins proposal for the non-abelian Born-Infeld action does work for
BPS states. Indeed, BPS states on $T^2$ correspond to abelian backgrounds
and on $T^4$ they are either abelian or self-dual. However, in this
subsection, we briefly comment on the situation on $T^6$.
Consider two D3-branes on $T^6$ alligned in the 146-hyperplane.
We rotate one of them over an angle $\gamma_1$ into the 12-plane, followed
by another rotation over an angle $\gamma_2$ in the 43-plane and finally
we rotate it by an angle $\gamma_3$ into the 65-plane. Such a
configuration should be described by choosing
constant magnetic backgrounds
${\cal F}_{2i-1\,2i}=f_i$ with $f_i=\tan (\gamma_i/2)$,
where $i\in\{1,2,3\}$. It seems impossible to express the
integrals in eqs. (\ref{l1}-\ref{l3}) in terms of known functions. However,
again we can perform the integrals by Taylor expanding all the
coefficients till a certain order around $f_i=0$. Doing this one notices
that once more, except when all backgroundfields are equal in
magnitude, the rescalings are such that the result cannot be brought
into a Yang-Mills form. When all background fields are equal in magnitude,
the Born-Infeld spectrum can be calculated from the corresponding
Yang-Mills spectrum. Once more, the Born-Infeld spectrum does not match the
spectrum obtained from a string theoretic calculation.
It looks highly unlikely to us
that the Born-Infeld spectrum will match the string spectrum for generic
choices of backgrounds/angles. This is reinforced by the fact that
through quartic order in the fieldstrength, the Born-Infeld action
scales correctly and does reproduce
correct results and this for arbitrary choices of the background fields.

If this is correct, this would imply that the non-abelian
Born-Infeld theory is not able to correctly describe BPS states on $T^6$.
In \cite{BDL}, the condition under which a configuration of branes at
angles preserves some supersymmetry was derived. With the above configuration
on $T^6$ this would correspond to D3-branes at angles
satisfying $\gamma_1 =
\gamma_2+\gamma_3$. In terms of the backgrounds, this is
\begin{eqnarray}
f_1=\frac{f_2+f_3}{1-f_2f_3}.
\end{eqnarray}

\setcounter{equation}{0}
\section{Multiply wrapped branes}

In this section, we consider a single multiply wrapped D-brane on
the torus $T^d$, with arbitrary wrapping structure. We give a
general construction of the fluctuation spectrum, discuss the
phenomenon of fractional momentum quantization, and explain how
this can be understood physically in terms of the string picture
of brane fluctuations. In this single brane case, the symmetrized
trace always reduces to an ordinary trace, and furthermore the
non-abelian Born-Infeld spectrum matches string theory
expectations. This indicates it is indeed the trace prescription
that needs to be reconsidered rather than the basic form of the
non-abelian Born-Infeld action.

We will study the most general case here, but it might be useful
to have in mind some concrete examples, like the class of
$D2$-brane wrappings on $T^2$ studied in \cite{HT}.

\subsection{Classification of brane wrappings}
\label{classification}

\begin{figure}
  \epsfig{file=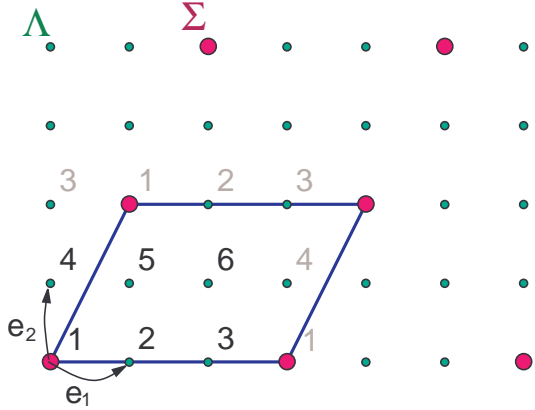,height=6cm,angle=0,trim = -170 0 0 0}
  \caption{A D2-brane wrapped 6 times around the torus $\IR^2 / \Lambda$
  can be represented as a quotient $\IR^2/\Sigma$ with $\Sigma$ a rank 6
  sublattice of $\Lambda$. For the lattice shown here,
  the classification parameters $m_{ij}$
  introduced in the text are $m_{11}=3$, $m_{22}=2$, $m_{21}=1$, $m_{12}=0$.
With the
  indicated sheet labeling $1,\ldots,6$, the closed path $e_1$ acts as the
  permutation $(123)(456)$ on the sheets (so $e_1[1]=2$, $e_1[2]=3$, $e_1[3]=1$
  etc.), while $e_2$ acts as $(143625)$. From this diagram, one can also read
off
  the winding numbers $\ell_i$ that appear further
  on in the momentum quantization condition: for example the off-diagonal
  mode generated by $E_{16}$ has $\ell_1=2, \ell_2=1$.}
  \label{fig-lattice}
\end{figure}

Consider an arbitrary p-brane wrapped (without branch points)
around the torus $T^p = \IR^p / \Lambda$, where $\Lambda = \left\{
(i_1 L_1,\ldots, i_p L_p) | i_1,\ldots,i_p \in \IZ \right\}$. The
D-brane, considered as a covering of the torus, can be identified
with $\IR^p / \Sigma$, where $\Sigma$ is a sublattice of $\Lambda$
(see fig \ref{fig-lattice} for an example). Different possible
sublattices correspond to different possible wrappings. Then $N$,
the number of times the brane covers the torus is equal to the
rank of $\Sigma$ in $\Lambda$, that is, the number of elements in
the group $\Lambda / \Sigma$.

The classification of N-fold wrappings is thus equivalent to the
classification of sublattices of rank N. A sublattice $\Sigma$ is
specified by a basis $\{s_i\}_i$ with $s_i = m_{ij} e_j$, where
$\{e_i\}_i$ is the standard basis of $\Lambda$ and $m_{ij}$ is an
integral matrix with determinant $N$. Bases that differ by an
$SL(p,\IZ)$ transformation are equivalent as they generate the
same lattice $\Sigma$. This equivalence can be used to transform
$m_{ij}$ into a lower triangular matrix, i.e. $m_{ij} = 0$ for
$i<j$. To see this, let us first consider the case $p=2$. An
$SL(2,\IZ)$ transformation acts on $m_{ij}$ as
\begin{equation}
  \left(
   \begin{array}{cc}
     m_{11} & m_{12} \\
     m_{21} & m_{22}
   \end{array}
  \right)
  \to
  \left(
   \begin{array}{cc}
     a & b \\
     c & d
   \end{array}
  \right)
  \left(
   \begin{array}{cc}
     m_{11} & m_{12} \\
     m_{21} & m_{22}
   \end{array}
  \right),
\end{equation}
so in particular $m_{12} \to a m_{12} + b m_{22}$. Choosing
$a=m_{22}/r$, $b=-m_{12}/r$ with $r=\mbox{gcd}(m_{12},m_{22})$,
and $c$, $d$ such that $d a - c b = 1$ we obtain the desired
$SL(2,\IZ)$ transformation putting $m_{12} = 0$. Similarly, for
$p>2$, $m_{ij}$ can be put in lower triangular form by repeated
application of elements of $SL(2,\IZ)$ subgroups of $SL(p,\IZ)$.

Now the residual $SL(p,\IZ)$ transformations are also lower
triangular, with all diagonal elements equal to $\pm 1$. The
latter can be taken to be $+1$ by putting all $m_{ii} > 0$ (we
assume $N>0$, i.e. the brane has positive orientation). The number
of times the brane wraps the $x^i$ direction\footnote{Note that
this ``winding number'' $m_i$ is not really canonically defined,
in the sense that it depends on how one chooses to gauge fix the
$SL(p,\IZ)$ equivalence.} is given by $m_i \equiv m_{ii}$, and $N
= \prod_i m_i$. The residual equivalence is fixed by requiring $0
\leq m_{ij} < m_{ii}$ for all $i,j$ with $i>j$. Indeed, for $p=2$,
the residual equivalence acts on $m_{ij}$ as $m_{21} \to m_{21} +
c \, m_{11}$, $c \in \IZ$ (with the other $m_{ij}$ unchanged), so
the equivalence is fixed by requiring $0 \leq m_{21} < m_{11}$.
For $p>2$, the analogous statement can be deduced by repeated
application of the $p=2$ reasoning.

So all in all we find
\begin{equation}
 w_{p,N} = \sum_{ \{ m_i \,:\, \prod_i m_i = N \} } (m_1)^{p-1} (m_2)^{p-2}
\cdots
 m_{p-1}
\end{equation}
inequivalent $N$-fold wrappings of $T^p$ by a single $p$-brane,
parametrized by $p$ winding numbers $m_i$ and $p(p-1)/2$ `mixing'
numbers $m_{ij}$, $i>j$.

\subsection{Implementation in worldvolume $U(N)$ gauge theory}

The low energy dynamics of a Dp-brane wrapped $N$ times around
$T^p$ is described by a $U(N)$ (Born-Infeld) gauge theory on $T^p$
with nontrivial boundary conditions ('t Hooft twists \cite{tH}),
depending on the structure of the wrapping, which as discussed
above is given by a sublattice $\Sigma$ of $\Lambda$. The $U(N)$
indices can be identified with the sheets of the covering
$\IR^p/\Sigma \to \IR^p/\Lambda$. In general, starting on a sheet
$a$ and moving along a closed path $\gamma$ in the torus, one ends
up on a different sheet $\gamma[a]$. The sheet permutation $a \to
\gamma[a]$ is (up to sheet relabeling) completely determined by
$\Sigma$. Indeed, a point on sheet $a$ of the D-brane with
coordinate $x$ on $\IR^p/\Lambda$ is represented by a unique point
$x + x(a)$ of the covering space $\IR^p/\Sigma$. Moving along a
closed loop $\gamma$ will move this to the point $x + x(a) +
x(\gamma) \mbox{ mod } \Sigma$ of $\IR/\Sigma$, which in turn
belongs to a certain sheet $\gamma[a]$, uniquely determined by
$\Sigma$ (and $\gamma$ of course). Conversely, $\Sigma$ is
uniquely determined by the permutations $a \to \gamma[a]$: it is
the lattice of points in the covering space $\IR^p$ reached by
paths $\sigma$ that start from $x=0$ and for which $\sigma[a] = a$
for all sheets $a$.

In the case of vanishing background fields, the wrapping structure
can be implemented by the following boundary conditions on fields
in the fundamental of $U(N)$:
\begin{eqnarray}
 \phi(L_1,x_2,\ldots,x_p) &=& \Omega_1 \, \phi(0,x_2,\ldots,x_p) \nonumber \\
 \phi(x_1,L_2,\ldots,x_p) &=& \Omega_2 \, \phi(x_1,0,\ldots,x_p) \nonumber \\
 &\cdots& \nonumber \\
 \phi(x_1,x_2,\ldots,L_p) &=& \Omega_p \, \phi(x_1,x_2,\ldots,0),
 \nonumber
\end{eqnarray}
where the $\Omega_i$ are the index permutation matrices
corresponding to the wrapping structure, that is, identifying the
basis element $e_i$ of $\Lambda$ with a closed loop in the $x^i$
direction:
\begin{equation}
{(\Omega_i)^a}_b = \delta_{e_i[a],b}
\end{equation}

An equivalent and more invariant description is in terms of
parallel transport. Consider an arbitrary path $\gamma$ between
points $P_i$ and $P_f$. Suppose $\gamma$ intersects the coordinate
boundary surfaces at the points $P_1, \ldots, P_n$. Then parallel
transport of $U(N)$ fundamentals along $\gamma$ is given by $\phi
\to U(\gamma) \phi$ with
\begin{equation}
 U(\gamma) \equiv \mbox{P} \exp [ i \int_{P_n}^{P_f} A ] \, \Omega(P_n)^{-1}
 \, \mbox{P} \exp [ i \int_{P_{n-1}}^{P_n} A ] \, \Omega(P_{n-1})^{-1} \cdots
 \, \mbox{P} \exp [ i \int_{P_i}^{P_1} A ],
\end{equation}
where the path ordered exponentials are over the indicated parts
of $\gamma$ and the $\Omega(P_i)$ are the appropriate $U(N)$
transition functions at the intersection points. On objects $X$
transforming in the adjoint, $U$ acts as $X \to U X U^{-1}$. Note
that by construction $U(\gamma_2 \circ \gamma_1) = U(\gamma_2)
U(\gamma_1)$.

Now in the case of zero background field ${\cal A}$, the
implementation of the wrapping structure in the gauge theory can
be described in terms of $U$ as
\begin{equation} \label{wrapcond}
 {U(\gamma)^a}_b = \delta_{\gamma[a],b}
\end{equation}
for any closed path $\gamma$. In particular for all $\sigma \in
\Sigma$ (that is paths starting at 0 and ending at a point of
$\Sigma$ when lifted to the covering space $\IR^p$), we have
\begin{equation}
 {U(\sigma)^a}_b = \delta_{a,b}
\end{equation}

When diagonal background fields are switched on, $U(\gamma)$ and
the boundary conditions will in general pick up additional
diagonal components (possibly path resp. position dependent),
changing (\ref{wrapcond}) in
\begin{equation} \label{Uab}
 {U(\gamma)^a}_b = \delta_{\gamma[a],b} \, e^{i \phi_b(\gamma)}.
\end{equation}
However, parallel transport of a diagonal matrix is insensitive to
the phase factors:
\begin{equation} \label{diagtransp}
 U(\gamma) \mbox{ diag}(d_a)_a \, U(\gamma)^{-1} = \mbox{
diag}(d_{\gamma[a]})_a
\end{equation}

Note that in general there exist many different equivalent gauge
theory descriptions of the same brane system, related by gauge
transformations $g(x)$ (not necessarily satisfying the boundary
conditions). Such transformations in general change both
background fields and boundary conditions. We introduced the
description in terms of $U(\gamma)$ here because, unlike boundary
conditions and background fields, it is invariant (up to
conjugation) under such transformations.

In this paper we only consider backgrounds with diagonal and
(covariant) constant field strength ${\cal F}$. This together with
(\ref{diagtransp}) and the fact that we are considering a single
wrapped brane implies that ${\cal F}$ is proportional to the unit
matrix $I_N$: ${\cal F}=F I_N$. Then for a closed contractible
path $C$, sweeping out the surface $S$ when contracted to a point,
we have
\begin{equation}
  U(C) = e^{i \int_S F} I_N,
\end{equation}
and for two closed paths $\gamma_1$, $\gamma_2$ with equal base
points:
\begin{equation} \label{commrel}
  U(\gamma_2) U(\gamma_1) = e^{i \int_{S(\gamma_1,\gamma_2)} F} \,
  U(\gamma_1) U(\gamma_2),
\end{equation}
where $S(\gamma_1,\gamma_2)$ is the surface swept out by
contracting $\gamma_2 \circ \gamma_1 \circ \gamma_2^{-1} \circ
\gamma_1^{-1}$. More explicitly, one has
\begin{equation} \label{fluxexpr}
\int_{S(\gamma_1,\gamma_2)} F = F_{ij} \int_{\gamma_1} dx^i \,
\int_{\gamma_2} dx^j.
\end{equation}
Since $U(\sigma)$ is diagonal for $\sigma \in \Sigma$ (cf.
equation (\ref{Uab})), the relation (\ref{commrel}) implies that
for $\sigma_1,\sigma_2 \in \Sigma$ we must have
\begin{equation}
  F_{ij} \int_{\sigma_1} dx^i \, \int_{\sigma_2} dx^j \in 2 \pi
  \IZ.
\end{equation}
Applying this to the basis elements $s_i=m_{ij} e_j$ of $\Sigma$
introduced in section 6.1,
yields the flux quantization condition
\begin{equation} \label{fluxquant}
  F_{ij} = \frac{2 \pi}{L_i L_j} (m^{-1})_{ik} \, (m^{-1})_{jl} \,
  q_{kl}
\end{equation}
where the $q_{kl}=-q_{lk}$ are integers.

Note also that parallel transport of objects transforming in the
adjoint is invariant under continuous path deformations.

\subsection{Fluctuation spectrum}

Because the field strength is proportional to the unit matrix, the
derivation of the fluctuation spectrum of the non-abelian
Born-Infled theory in this background is quite analogous to the
derivation in the abelian $U(1)$ case discussed in section 2.
The main difference is the modified quantization condition on the
momenta. When the brane carries flux, this modification is
nontrivial \cite{HT}. Below we give a general construction of the
spectrum, independent of specific background field and boundary
condition choices, which clearly shows the physical origin of the
modifications to the momentum quantization condition.

The fluctuation modes of the quadratic Lagrangian (\ref{int1}), in
the gauge $A_0=0$, $D_i A^i = 0$, are of the form $\delta A_i =
u_i \delta A$ with $k^i u_i =0$ and $\delta A$ a solution of
\begin{equation} \label{modeq}
 D_j \delta A = i k_j \delta A,
\end{equation}
with energy
\begin{equation} \label{flucspec}
 M^2 = {\cal G}^{ij} k_i k_j
\end{equation}
where ${\cal G}$ is given by (\ref{defg}) as usual. Solutions to
(\ref{modeq}) are of the form
\begin{equation} \label{gensol}
 \delta A(x) = U(\gamma_x) \, \delta A_0 \, U(\gamma_x)^{-1} \, e^{i
\int_{\gamma_x} k_i dx^i},
\end{equation}
where $\gamma_x$ is any path from 0 to $x$. Because $U$ acting in
the adjoint is invariant under continuous path deformations,
(\ref{gensol}) gives indeed a well defined globally smooth
solution, provided
\begin{equation} \label{globalcond}
 \delta A_0 = U(\lambda) \, \delta A_0 \, U(\lambda)^{-1}
 \, e^{i \int_\lambda k_i dx^i}
\end{equation}
for any closed path $\lambda$ based at $0$. Such paths can be
identified with the points of the lattice $\Lambda$, hence the
notation $\lambda$. Note that if condition (\ref{globalcond}) is
satisfied, one trivially has
\begin{equation} \label{average}
 \delta A_0 = \lim_{\Lambda^\prime \to \Lambda} \frac{1}{|\Lambda^\prime|}
 \sum_{\lambda^\prime \in \Lambda^\prime} U(\lambda^\prime) \, \delta
 A_0\,
 U(\lambda^\prime)^{-1} \, e^{i \int_{\lambda^\prime} k_i dx^i},
\end{equation}
where $\Lambda^\prime$ denotes a subset of $\Lambda$ of size
$|\Lambda^\prime|$. Conversely, any $\delta A_0$ {\em defined} by
\begin{equation} \label{defA0}
 \delta A_0
 \equiv \langle E \rangle_k
 \equiv \lim_{\Lambda^\prime \to \Lambda} \frac{1}{|\Lambda^\prime|}
 \sum_{\lambda^\prime \in \Lambda^\prime} U(\lambda^\prime) \,
 E \, U(\lambda^\prime)^{-1} \, e^{i \int_{\lambda^\prime} k_i dx^i},
\end{equation}
with $E$ an arbitrary matrix, automatically satisfies condition
(\ref{globalcond}), as follows from $U(\lambda) U(\lambda^\prime)
= U(\lambda \circ \lambda^\prime)$ and invariance up to a phase of
(\ref{average}) under the shift $\lambda^\prime \to \lambda \circ
\lambda^\prime$.

So the general solution to (\ref{globalcond}) is a linear
combination of the matrices $\langle E_{ab} \rangle_k$, with the
matrix basis $E_{ab}$, $a,b=1,\ldots,N $ as defined under equation
(\ref{Eab}). Now $\langle E_{ab} \rangle_k$ is only nonzero for
specific values of the momentum $k$. To see this, first note that
(\ref{Uab}) implies we can write $E_{ab}=U(\gamma_{ab}) D_b$ with
$\gamma_{ab}$ a closed path that, lifted to the covering space,
runs from 0 lifted to sheet $a$ to 0 lifted to sheet $b$, so
$\gamma_{ab}[a] = b$, and with $D_b$ the diagonal matrix defined
by $(D_b)_{cc} = e^{-i \phi_b(\gamma_{ab})} \delta_{cb}$. Then for
$\sigma \in \Sigma$, we have
\begin{eqnarray}
  \langle E_{ab} \rangle_k &=& \langle U(\gamma_{ab}) D_b \rangle_k \nonumber
\\
  &=& \exp [i \int_\sigma k_i dx^i] \, \langle U(\sigma) U(\gamma_{ab}) \, D_b
\, U(\sigma)^{-1} \rangle_k  \nonumber \\
  &=& \exp [i \int_\sigma k_i dx^i] \, \langle \exp[i
\int_{S(\gamma_{ab},\sigma)} F] \,
   U(\gamma_{ab}) U(\sigma) \, D_b \, U(\sigma)^{-1} \rangle_k \nonumber \\
  &=& \exp[i \int_\sigma k_i dx^i + i \int_{S(\gamma_{ab},\sigma)} F] \,
\langle U(\gamma_{ab}) \, D_b \, \rangle_k \nonumber \\
  &=& \exp[i \int_\sigma k_i dx^i + i \int_{S(\gamma_{ab},\sigma)} F] \,
\langle E_{ab}
   \rangle_k. \nonumber
\end{eqnarray}
Equation (\ref{commrel}) was used in going from the second to the
third line, and (\ref{diagtransp}) together with $\sigma[b]=b$ to
go from the third line to the fourth. So $\langle E_{ab}
\rangle_k$ can only be nonzero if the following momentum
quantization condition is satisfied:
\begin{equation} \label{mqc}
  \int_\sigma k_i dx^i + \int_{S(\gamma_{ab},\sigma)} F \in 2 \pi \IZ
\end{equation}
for all $\sigma \in \Sigma$. For diagonal excitations, the flux
term is zero and the quantization condition is simply the usual
one for a particle on a compact space --- here the covering space
$\IR^p/\Sigma$ corresponding to the wrapped brane, as could be
expected. However, off-diagonal excitations pick up an additional
phase when going around a loop, and the quantization condition is
shifted. In the string picture of these excitations, where the
off-diagonal fluctuations correspond to open strings with
endpoints on different sheets, this additional term is easily
understood: it is nothing but the usual coupling of a string
running from sheet $a$ to sheet $b$ and going around a loop
$\sigma$ of the brane, to the given gauge field background. Also
the effective metric ${\cal G}$ appearing in (\ref{flucspec}) can
be understood directly in the string theory picture, see e.g.
\cite{SWnc}. So here the non-abelian Born-Infeld theory matches
nicely with string theory expectations, and in particular we
expect the fluctuation spectrum to be reproduced by string theory
on an appropriate T-dual system, as in \cite{HT}, though we did
not work out the details for the general case.

Let us make (\ref{mqc}) more explicit. Using (\ref{fluxexpr}) and
(\ref{fluxquant}), and taking $\sigma$ equal to the basis elements
$s_i=m_{ij} e_j$ introduced in section 6.1,
the quantization condition can be rewritten as
\begin{equation} \label{qcexpl}
  k_i = \frac{2 \pi}{L_i} (m^{-1})_{ij} \, [ n_j
  + (m^{-1})_{kl} \, q_{jl} \, \ell_k ],
\end{equation}
where the $n_j$ are arbitrary integers, $q_{jl}=-q_{lj}$ are the
flux quantum numbers from (\ref{fluxquant}), and $\ell_k$ is the
winding number of $\gamma_{ab}$ in the $x^k$ direction:
$\gamma_{ab}=l_k e_k$. The precise fractionalization of momentum
depends on the details of the different integers involved, but the
minimal momentum quantum will never be less than $1/N$.

Note that if the momentum quantization condition is satisfied, the
infinite sum in (\ref{defA0}) can be reduced to a finite sum over
$\Lambda/\Sigma$:
\begin{equation}
 \langle E_{ab} \rangle_k
 = \frac{1}{N}
 \sum_{\lambda^\prime \in \Lambda/\Sigma} U(\lambda^\prime) \,
 E_{ab} \, U(\lambda^\prime)^{-1} \, e^{i \int_{\lambda^\prime} k_i dx^i},
\end{equation}
because paths $\lambda^\prime$ and $\lambda^\prime \circ \sigma$,
with $\sigma \in \Sigma$, give the same contribution. This gives
an in principle straightforward way to compute the $\langle E_{ab}
\rangle_k$ explicitly for a given $U(N)$ bundle. It also shows
that all modes constructed from $\langle E_{ab} \rangle_k$ are
nonzero, because all terms in the finite sum are linearly
independent matrices. Finally note that all $\langle E_{ab}
\rangle_k$ with the same $\gamma_{ab}$ (i.e. the same $\ell_i$)
are actually equal up to a phase factor, as they are mapped to
each other by a path shift $\lambda^\prime \to \lambda^\prime
\circ \gamma$ for a certain $\gamma$. This corresponds to the fact
that only the relative position (in the covering space) of the
sheets is physical. So the quantum numbers $n_i$ and $\ell_i$
appearing in (\ref{qcexpl}) label the modes without degeneracy.

\setcounter{equation}{0}
\section{Conclusions}
The calculation of the mass spectrum is probably the simplest question
which can be addressed by means of the effective action. It only probes
the effective action through second order in the fluctuations. As we
demonstrated in this paper, the non-abelian Born-Infeld action defined
using the symmetrized trace prescription is not able to reproduce the
spectrum as predicted by string theory. Not surprisingly, the
first two terms of the Born-Infeld action, the $F^2$ and $F^4$ terms,
give correct answers. However at order $F^6$ and higher, things go wrong.
Keeping in mind that the non-abelian Born-Infeld action should reduce to
the sum of abelian Born-infeld actions for well separated D-branes,
additional corrections should involve commutators of the field strengths.
As advocated by Tseytlin, this can be understood as follows. At higher
order one expects corrections going as derivatives acting on
the fieldstrength \cite{der} which, under the assumption that the
velocities vary slowly, are ignored in the effective action. However this
is ambiguous in a non-abelian gauge theory as one has that
\begin{eqnarray}
D_\alpha D_\beta F_{\gamma\delta}=\frac 1 2 \{D_\alpha ,
D_\beta\}F_{\gamma\delta}-\frac i 2 {[}F_{\alpha \beta}, F_{\gamma\delta}
{]}.
\end{eqnarray}
It is clear that the symmetrized product of
derivatives acting on a fieldstrength should be viewed as an acceleration
term which can safely be neglected. The anti-symmetrized products
however should be kept and will contribute to the mass spectrum!
Precisely these terms are not captured by the proposal in \cite{AT}.

The present work clearly demonstrates the need to address this problem.
One possible way would be to construct these terms using string
field theory along the lines developed in \cite{WT2}.
However in order to make concrete statements about the $F^6$ terms, the
calculation has to be pushed to an order which is probably unfeasible
without the aid of a computer. Another way to get the $F^6$ terms would be by
calculating a five loop beta function. Due to the fact that for the present
purpose it is sufficient to consider trivial gravitational backgrounds,
the vertices appearing are rather simple. Despite of this, the calculation
remains very involved. Yet another approach could consist of supersymmetrizing
the non-abelian Born-Infeld action. It might be that the requirement of
invariance under $\kappa$-symmetry favorizes certain ordenings.

Perhaps the simplest way to proceed is by using the mass spectrum as a
guideline. As a first step, one would have to make a systematic
analysis of possible modifications of the $F^6$ terms in the non-abelian
Born-Infeld action. In \cite{B}, a first attempt was already made but it
was clear that the $T^4$ case was not sufficient to unambigously
determine the $F^6$ term. Now, many more examples are available. A priori
we expect 5! different permutations of the Lie algebraic factors to be
relevant. However, as the mass spectrum automatically narrows the group
to a $U(2)$ factor, the number of truly inequivalent permutations is
considerably smaller. In fact a closer analysis shows that there are
only 15 possibilities. Furthermore, ignoring the
Lorentz indices all together, only 5 inequivalent ordenings remain.
The two-torus, being a particularly simple case, is the best place to
start the analysis.

Additional hints are provided by the BPS configurations on $T^4$ for which
Tseytlins prescription does provide the correct spectrum. On $T^6$ and
$T^8$ further insights are provided by the fact that in the Yang-Mills
case BPS configurations involve linear relations between the backgrounds
\cite{JT}, while for the Born-Infeld theory the backgrounds should
satisfy non-linear relations \cite{BDL}. Indeed, focussing on $T^6$ we saw
that at the level of the Yang-Mills theory the BPS relation is given by
$f_1-f_2-f_3=0$ while in the Born-Infeld theory it becomes
\begin{eqnarray}
f_1-f_2-f_3=(2\pi\alpha ')^2 f_1 f_2 f_3.
\end{eqnarray}
We reinserted the factors of $2\pi\alpha '$ in order to clearly
demonstrate that this indeed relates different orders in $F$ in the
Born-Infeld theory.

We also know that the
Born-Infeld spectrum should have the same form as the Yang-Mills spectrum,
but with rescaled backgrounds. As became clear from section 5, this puts
strong additional constraints on the possible modifications.
Finally, requiring a consistent behaviour of the Born-Infeld action under
T-duality further reduces the possibilities. A systematic
study of modifications of the non-abelian Born-Infeld action satisfying
all requirements will be presented in a separate publication
\cite{UW}.

The initial exploration of general wrappings of D-branes on
tori, opens several interesting questions, such as the precise
implications of T-duality transformations and D-brane configurations
on more involved geometries, which deserve further
investigation. We hope to return to this in the future.

\vspace{5mm}

\noindent {\bf Acknowledgments}:
We would like to thank Walter Troost for several illuminating discussions
and Alberto Santambrogio for a useful comment.
A.S. and J.T. are supported in part by the FWO and by the European
Commission TMR programme ERBFMRX-CT96-0045 in which both are
associated to K.\ U.\ Leuven.

\end{document}